\documentclass[aps,prl,twocolumn,groupedaddress,showpacs,preprintnumbers,superscriptaddress]{revtex4}
\usepackage{epsfig}
\usepackage{latexsym}
\usepackage{graphicx}
\usepackage{bm}

\begin{document}

\bibliographystyle{apsrev}

\preprint{Edinburgh 2007/44, KEK-TH-1192, RBRC-693}

\title{
Nucleon axial charge in 2+1 flavor dynamical lattice QCD 
with domain wall fermions}

\author{T.~Yamazaki}
\affiliation{Physics Department, University of Connecticut, Storrs CT, 06269-3046, USA}
\author{Y.~Aoki}
\affiliation{RIKEN-BNL Research Center, Brookhaven National Laboratory, Upton, NY 11973, USA}
\author{T.~Blum}
\affiliation{Physics Department, University of Connecticut, Storrs CT, 06269-3046, USA}
\affiliation{RIKEN-BNL Research Center, Brookhaven National Laboratory, Upton, NY 11973, USA}
\author{H.~W.~Lin}
\affiliation{Thomas Jefferson National Accelerator Facility, Newport News, VA 23606, USA}
\author{M.~F.~Lin}
\affiliation{Center for Theoretical Physics, Massachusetts Institute of Technology, Cambridge, MA 02139, USA}
\author{S.~Ohta}
\affiliation{Institute of Particle and Nuclear Studies, KEK, Tsukuba, 305-0801, Japan}
\affiliation{Physics Department, Sokendai Graduate U.\ Adv.\ Studies, Hayama, Kanagawa 240-0193, Japan}
\affiliation{RIKEN-BNL Research Center, Brookhaven National Laboratory, Upton, NY 11973, USA}
\author{S.~Sasaki}
\affiliation{Department of Physics, University of Tokyo, Hongo 7-3-1, Bunkyo-ku, Tokyo 113, Japan}
\author{R.~J.~Tweedie}
\affiliation{School of Physics, The University of Edinburgh, Edinburgh EH9 3JZ, UK}
\author{J.~M.~Zanotti}
\affiliation{School of Physics, The University of Edinburgh, Edinburgh EH9 3JZ, UK}
\collaboration{RBC+UKQCD Collaborations}

\date{
\today
}

\begin{abstract}
We present results for the nucleon axial charge $g_A$ at a fixed lattice
spacing of $1/a=1.73(3)$ GeV using 2+1 flavors of domain wall fermions
on size $16^3\times 32$ and $24^3\times 64$ lattices ($L=1.8$ and 2.7 fm)
with length 16 in the fifth dimension.
The length of the Monte Carlo trajectory at the lightest $m_\pi$
is 7360 units, including 900 for thermalization.
We find finite volume effects are larger
than the pion mass dependence at $m_\pi= 330$ MeV. We also find
that $g_A$ exhibits a scaling with the single variable 
$m_\pi L$ which 
can also be seen in previous two-flavor domain wall and Wilson fermion calculations.
Using this scaling to eliminate the finite-volume effect, 
we obtain $g_A = 1.20(6)(4)$ at the physical pion mass, 
$m_\pi = 135$ MeV, where the first and second errors are statistical
and systematic. 
The observed finite-volume scaling also appears in similar quenched 
simulations, but disappear when $V\ge (2.4$ fm)$^3$. 
We argue this is a dynamical quark effect.
\end{abstract}

\maketitle

The isovector axial charge
of nucleon is a fundamental observable in hadron physics.
It is defined as the axial vector form factor
at zero four-momentum transfer, $g_A = G_A(0)$.
The axial vector form factor is given by
the nucleon matrix element of the axial vector current,
$A^a_\mu = \overline{\psi}\gamma_\mu\gamma_5 (\tau^a/2) \psi$,
with $u,d$ quark doublet $\psi$,
$\displaystyle{
\langle n^\prime | A^a_\mu | n \rangle =
\overline{u}_{n^\prime} [ \gamma_\mu G_A(q^2) 
+ i q_\mu G_P(q^2) ] \gamma_5 (\tau^a /2) u_n,
}$
where $G_P$ is the induced pseudoscalar form factor,
$\tau^a$ an isospin Pauli matrix, and $q$ the momentum
transfer, $q_\mu = p^n_\mu - p^{n^\prime}_\mu$.
Experimentaly $g_A$ has been obtained very precisely, 
$g_A=1.2695(29)$~\cite{PDBook}, 
through neutron $\beta$ decay.

$g_A$ is related to 
the spontaneous breaking of the chiral symmetry of the strong interaction
through the well-known Goldberger-Treiman relation~\cite{Goldberger:1958tr}.
This relation shows that $g_A$ is proportional to 
the strong pion-nucleon coupling at $q^2\approx 0$.
Furthermore the Adler-Weisberger sum rule~\cite{Weisberger:1965hp,Adler:1965ka}
reveals that $g_A$ differs from unity
for a structure-less nucleon through 
the difference between the integrals of the total cross sections of 
the $\pi^+ p$ and $\pi^- p$ channels.
These are in good agreement with experiments.

Hence the axial charge allows us to perform a precision test of
(lattice) QCD in the baryon sector. Moreover, since it is an
isovector matrix element, only connected quark diagrams at $q^2=0$
contribute, making the calculation technically simpler. Furthermore, the
renormalization of the axial vector current is simplified when
utilizing a lattice chiral fermion action such as 
the domain wall fermion (DWF) action~\cite{Kaplan:1992bt,Shamir:1993zy,Furman:1994ky}.
However, it is difficult to control finite-volume effects (FVE's)
in $g_A$, as suggested by quenched 
calculations~\cite{Sasaki:2003jh}.
The FVE's have been investigated in 
effective models~\cite{Jaffe:2001eb,Thomas:2005qm},
and also in heavy baryon chiral perturbation theory 
(HBChPT)~\cite{Beane:2004rf,Detmold:2005pt,Khan:2006de}. 
The FVE's in HBChPT are
inconsistent with lattice calculations
unless contributions of the $\Delta$ baryon resonance are included.
Such contributions introduce several unknown parameters, and the FVE
seems sensitive to one of them, the $\Delta$-$N$ 
coupling~\cite{Beane:2004rf,Khan:2006de,Lin:2008uz}.
Until now there has been only one investigation in the light 
quark mass region
with 2+1 flavors, the mixed action calculation~\cite{Edwards:2005ym}
by LHPC
which reported the FVE is small for
a volume of $V =($2.5 fm)$^3$.

In this letter we report a result for $g_A$
using 2+1 flavors of dynamical DWF 
with several quark masses corresponding to $m_\pi = 0.33$--0.67 GeV, 
on two volumes with spatial size $L=1.8$ and 2.7 fm.
Our light quark masses and the spatial volumes allow
a detailed investigation of the FVE
in full QCD
where the sea and valence quarks are identical. It is found that the FVE is not negligible
at the lightest $m_\pi$, even on the larger volume, and 
that $g_A$ exhibits a scaling behavior with 
the single variable $m_\pi L$.

%
Our calculation is performed with a fixed lattice spacing, two
space-time lattice sizes, $16^3 \times 32$ and $24^3 \times 64$,
the Iwasaki gauge action~\cite{Iwasaki:1984cj} with $\beta=2.13$,
and the DWF action 
with a fifth dimension of size $16$ and $M_5 = 1.8$. 
Each ensemble of configurations uses the same dynamical strange quark mass, 
$m_s^{\rm sea} = 0.04$ in lattice units.  We use
four light sea quark masses on the $24^3$ lattice, $m_{\rm sea} = 0.005, 0.01, 0.02,$ and $0.03$ and the same heaviest
three masses on the $16^3$ lattice.
The ensembles, described in \cite{Allton:2007hx,Boyle:2007mk},
were generated using the RHMC algorithm~\cite{Clark:2006fx} with 
trajectories of unit length.  
The measurements were performed at the unitary points only,
$m_f = m_{\rm val}=m_{\rm sea}$.
We use the mass of the $\Omega^-$ baryon to determine the inverse of the lattice spacing 
$1/a=1.73(3)$ GeV~\cite{Lin:2007pt,RBC-UKQCD:2007}. 
The residual quark mass due to the finite size of the fifth dimension
is $0.00315(2)$.
The non-zero lattice spacing error is small in our calculation
because the DWF action is automatically off-shell $O(a)$ improved.

Four measurements are carried out for the $24^3$ ensembles on 
each configuration.
The number of Monte Carlo trajectories used for measurements is 6460, 3560, 2000, and 2120 for $m_f = 0.005$,
0.01, 0.02, 0.03, respectively, with 10 trajectory separations 
for $m_f = 0.005$, 0.01 and 20 for 0.02, 0.03.
The measurements are blocked into bins of 40 trajectories each to reduce auto-correlations. On the $16^3$ ensembles we use 
3500 trajectories separated by 
10 trajectories at $m_f = 0.01$, and 0.03, and by 5 at 0.02.
The data are blocked with 20 trajectories per bin.

The axial charge is calculated from the ratio of the matrix elements
of the spatial component of the axial vector current and
the temporal component of 
the vector current, $V^a_t = \overline{\psi}\gamma_t (\tau^a/2)\psi$,
$\displaystyle{
\langle n^\prime | A^a_i | n \rangle / 
\langle n^\prime | V^a_t | n \rangle
= g_A.
}$
This ratio gives the renormalized axial charge because
$A_\mu$ and $V_\mu$ share a common renormalization constant
due to the chiral symmetry of DWF.
In our simulation
the two constants are consistent to less than 0.5\% at the chiral limit.
In order to increase the overlap with the ground state, the quark
propagators are calculated with gauge invariant Gaussian 
smearing~\cite{Alexandrou:1992ti} and
we employ sufficient separation in Euclidean time, 
more than 1.37 fm, which is the largest
used so far in dynamical calculations
of $g_A$~\cite{Edwards:2005ym,Dolgov:2002zm,Alexandrou:2007xj}, 
between the location
of the nucleon source and sink to minimize excited
state contamination.

\begin{figure}
\includegraphics*[angle=0,width=0.48\textwidth]{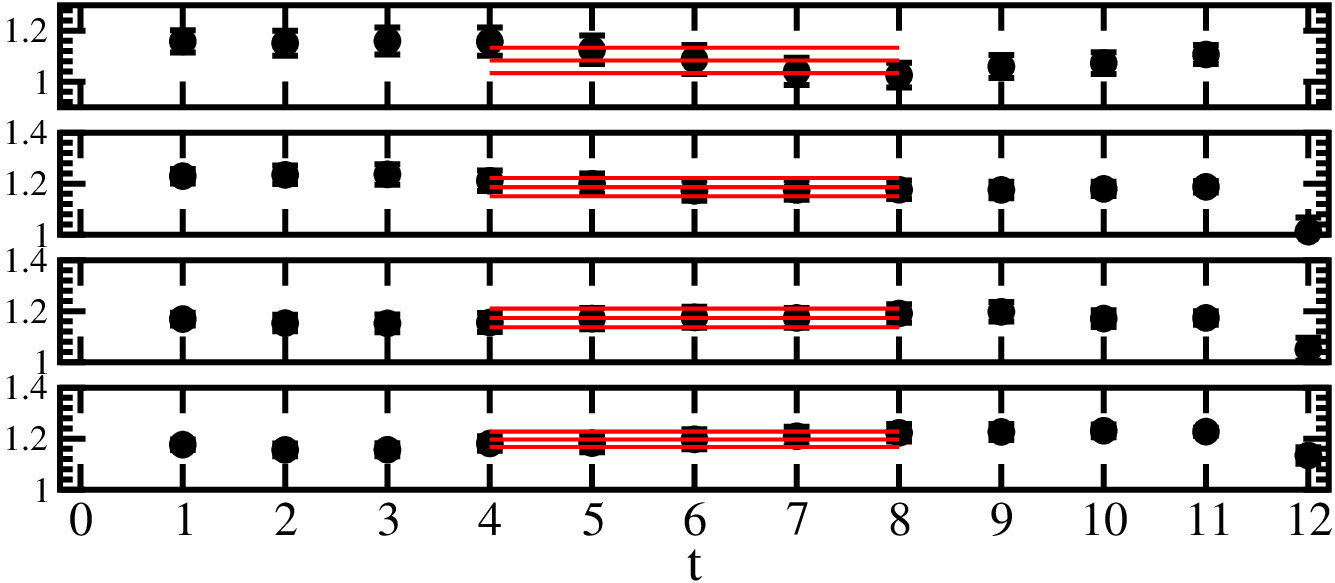}
\vspace{-0.3in}
\caption{\label{fig:plateau}
Plateaus of $g_A$. $V=(2.7$ fm$)^3$ and $m_f = 0.005$, 0.01, 0.02, and 0.03,
from top to bottom.
}
\vspace{-0.1in}
\end{figure}

The plateaus of $g_A$ computed on volume $V=(2.7$ fm$)^3$ are shown 
in Fig.~\ref{fig:plateau}. 
We checked that consistent results are obtained 
by either fitting or averaging over appropriate time slices,
$t=4$--8, and also by fitting the data symmetrized about $t=6$.
The larger volume data can be symmetrized
because the source and sink operators are identical in the limit of large statistics.
We note that the length of our lightest mass run is already the longest we know of 
for comparable simulation parameters. 
Results obtained from the fit using the
unsymmetrized data, presented in the figure with
one standard deviation, are employed in the analysis.

%

%

%
\begin{figure}
\includegraphics*[angle=0,width=0.48\textwidth]{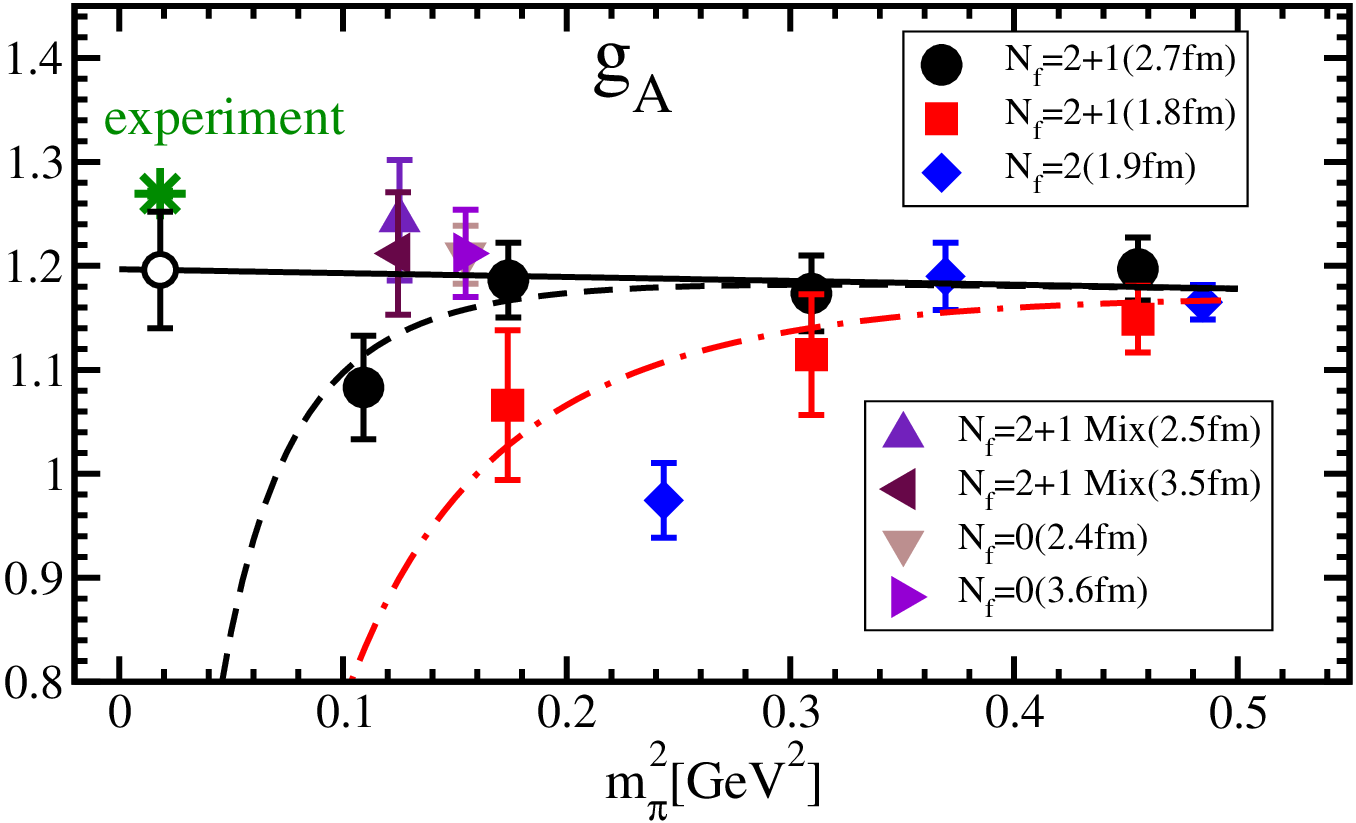}
\vspace{-0.3in}
\caption{\label{fig:gagv_mpi}
$g_A$.
Dashed and solid lines denote the
fit results and chiral extrapolation in infinite volume, respectively.
The open circle is extrapolated result at $m_\pi = 135$ MeV.
}
\vspace{-0.1in}
\end{figure}
\begin{table}[t!]
\begin{tabular}{ccccc}\hline\hline
$m_f$ & 0.005 & 0.01 & 0.02 & 0.03 \\\hline
$m_\pi$[GeV]\cite{RBC-UKQCD:2007} & 
0.3313(13) & 0.4189(13) & 0.5572(5) & 0.6721(6) \\
(2.7 fm)$^3$ & 1.083(50) & 1.186(36) & 1.173(36) & 1.197(30) \\
(1.8 fm)$^3$ & N/A       & 1.066(72) & 1.115(58) & 1.149(32) \\\hline\hline
\end{tabular}
\vspace{-0.1in}
\caption{\label{tab:gagv} $g_A$ and $m_\pi$ ($V=(2.7$ fm$)^3$ only). }
\vspace{-0.2in}
\end{table}

Figure~\ref{fig:gagv_mpi} shows our result for $g_A$.
The results are also presented in Table~\ref{tab:gagv}.
The (2.7 fm)$^3$ data are almost independent of the pion mass (squared) except for the lightest point
which is about 9\% smaller than the others. 
A set of the results obtained with a smaller volume, (1.8 fm)$^3$
shows a similar downward behavior,
albeit with relatively larger statistical uncertainties.
An earlier two flavor calculation by 
RBC~\cite{Lin:2008uz} with spatial volume (1.9 fm)$^3$ 
and $1/a=1.7$ GeV showed a clear downward behavior, 
but it sets in at heavier pion mass.

We suspect that this pion mass dependence driving $g_A$ away from
the experimental value is caused by the finite 
volume of our calculation: in general such an effect is expected to grow as
the quark mass gets lighter at fixed volume, or the volume decreases for fixed quark mass. 
More quantitatively, we observe in the figure
that the two flavor result with $V=(1.9$ fm)$^3$
significantly decreases at $m_\pi^2 \approx 0.24$ GeV$^2$, while the 2+1 flavor
results with $V=(2.7$ fm)$^3$ do not decrease even at $m_\pi^2 \approx 0.17$ GeV$^2$.
Another trend of the FVE seen in Fig.~\ref{fig:gagv_mpi}
is that all the 2+1 flavor, smaller volume data are systematically lower than 
the larger volume data.
Similar behavior was observed in
quenched DWF studies~\cite{Sasaki:2003jh,Sasaki:2007gw}.
However, for pion masses close to our lightest point
such a sizable shift
is not observed when $V$ is larger than about (2.4 fm)$^3$,
not only in the quenched case,
but also the 2+1 flavor, mixed action, calculation in~\cite{Edwards:2005ym},
as shown in Fig.~\ref{fig:gagv_mpi}. On the other hand,
our results suggest that $V=(2.7$ fm)$^3$ is not enough to
avoid a significant FVE on $g_A$
when $m_\pi \le 0.33$ GeV in dynamical fermion calculations.

\begin{figure}[t]
\includegraphics*[angle=0,width=0.48\textwidth]{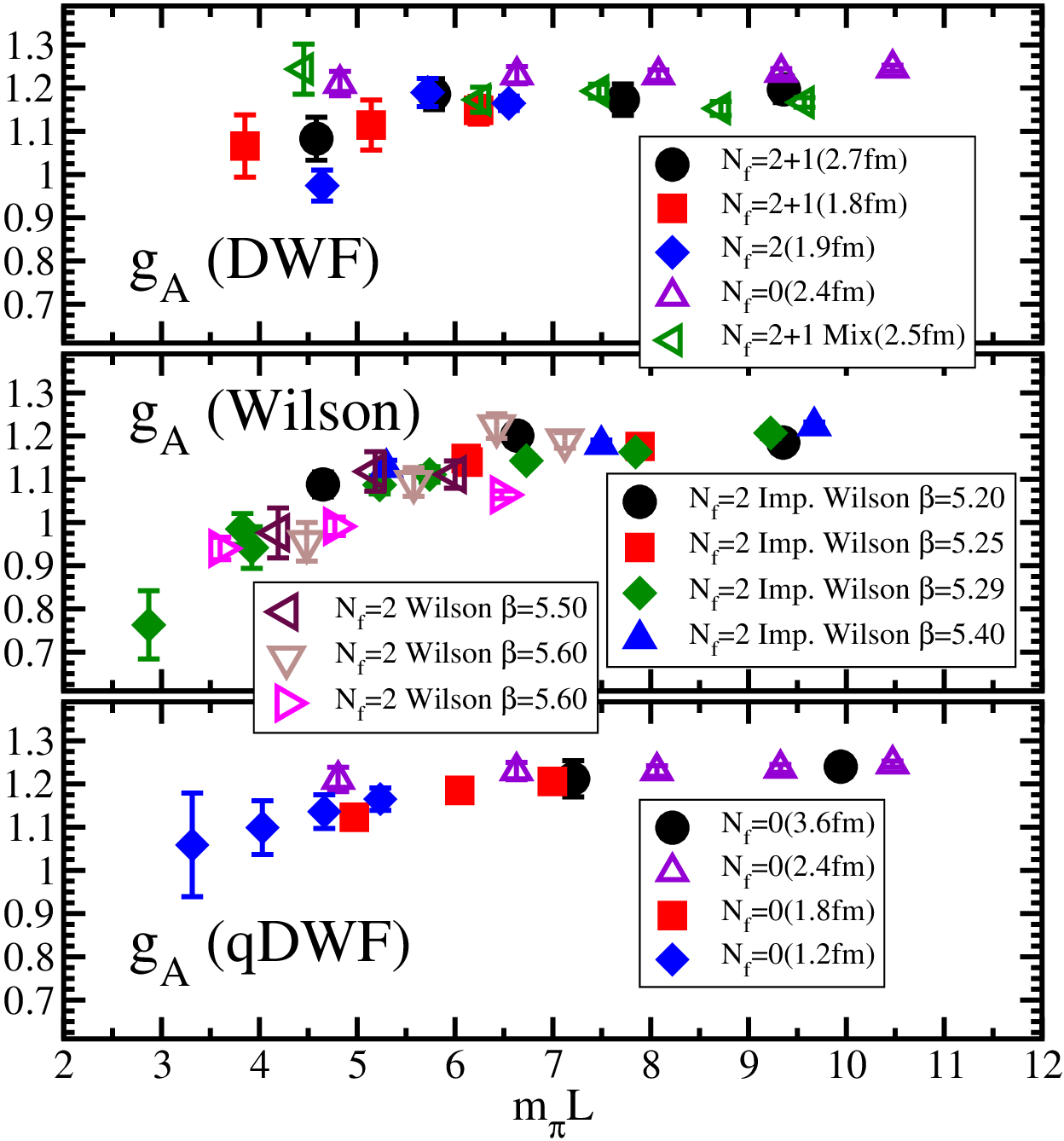}
\vspace{-0.3in}
\caption{\label{fig:gagv_mpiL}
$m_\pi L$ scaling of $g_A$.
Top, middle, and bottom panels are dynamical and mixed action DWF, 
dynamical Wilson,
and quenched DWF results, respectively.
In bottom panel, open symbol is same as in top panel.
}
\vspace{-0.2in}
\end{figure}

In order to more directly compare the various results, 
we plot $g_A$ against a dimensionless quantity, $m_\pi L$,
in the top panel of Fig.~\ref{fig:gagv_mpiL}.
Interestingly, we find that the 2+1 flavor results on both volumes and the two flavor ones
reasonably collapse onto a single curve that monotonically increases with $m_\pi L$;  in other words, they exhibit scaling in $m_\pi L$. 
A similar scaling also appears in dynamical two flavor 
(improved) Wilson fermion calculations
as shown in the middle panel of  Fig.~\ref{fig:gagv_mpiL}~\cite{Khan:2006de,Dolgov:2002zm,Alexandrou:2007xj} for
the unitary points $\kappa_{\rm sea} = \kappa_{\rm val}$,
with various volumes (0.95--2.0 fm)$^3$,
pion masses 0.38--1.18 GeV, and gauge couplings.
The large difference in Wilson data at $\beta = 5.60$ on $m_\pi L \sim 6.5$
can be described by different choices of the renormalization constant
of $A_\mu$.
While the trend is similar in the quenched DWF case~\cite{Sasaki:2003jh,Sasaki:2007gw} with pion masses in the range 0.39--0.86 GeV
and $1/a= 1.3$ GeV (see bottom panel, Fig.~\ref{fig:gagv_mpiL}),
the scaling is violated for the point with smallest $m_\pi L$
on $V=(2.4$ fm)$^3$.
The lightest point does not follow the (1.8 fm)$^3$ 
data: they differ by 2.5 standard deviations ($\sigma$)
at $m_\pi L \sim 5$, suggesting that there are non-universal terms that
depend separately on $m_\pi$ and $V$. In particular, this effect may be
due to the presence of a quenched chiral log~\cite{Kim:1996bz}.
From~\cite{Kim:1996bz}, the size of the effect at this mass can readily explain
the discrepancy observed with the dynamical $m_\pi L$ scaling.
Note, at this mass,
but going to $V=(3.6$ fm)$^3$, no FVE is detected in 
the quenched case as can be seen in Fig.~\ref{fig:gagv_mpi}.

It is interesting to compare our larger volume, 2+1 flavor result with
the mixed action, 2+1 flavor result with a similar volume~\cite{Edwards:2005ym},
denoted by the left triangle
in the top panel of Fig.~\ref{fig:gagv_mpiL}.
At heavy pion mass the results are statistically
consistent and essentially independent of $m_\pi L$.
At $m_\pi L \sim 4.5$ the mixed action result, however, is larger than
ours by (a combined) 2.1$\sigma$,
but consistent with the quenched DWF result with (2.4 fm)$^3$ 
volume~\cite{Sasaki:2003jh}, (the up triangle
in the figure),
which is also 2.2$\sigma$ larger than our result. The mixed action result was also calculated with an even larger volume, and no
FVE was detected~\cite{Edwards:2005ym}. 
Again, similar to the quenched case.

A possible explanation of the difference is that it is simply a dynamical fermion effect.
The mixed action results are partially quenched, calculated 
with improved staggered sea quarks and domain wall valence quarks.
The pion mass of valence DWF is tuned
to match the lightest pseudoscalar meson mass computed entirely with staggered fermions.
However, there is an ambiguity~\cite{Bar:2005tu,Prelovsek:2005rf} 
in choosing the staggered pseudoscalar meson for this tuning since there are several.
Thus, if the valence quark is effectively much lighter than the 
sea quark,
a mixed action calculation effectively becomes quenched.
This may lead to a deviation from the unitary calculation
and consistency with the quenched calculation.
Mixed action ChPT 
shows the presence of partially-quenched logs whose size is
consistent with the observed effect~\cite{Jiang:2007sn,Chen:2007ug}.

%

For the chiral extrapolation of $g_A$,
we attempt to include the FVE in our data.
While the pion mass dependence of $g_A$, including the FVE, 
has been investigated in the small scale expansion
(SSE) scheme of HBChPT~\cite{Khan:2006de}, 
the size of the FVE
on $V=(2.7$ fm)$^3$ predicted in SSE
is less than 1\% in our pion mass region. The correction is much too small
to account for the observed FVE in our data.
This suggests that the FVE in HBChPT, which
is estimated by replacing all loop integrals by summations,
is not the leading FVE in $g_A$, as one in $\varepsilon$ 
regime~\cite{Smigielski:2007pe}.
We also note that our attempts
to fit the mass dependence of the data to HBChPT failed, which is likely due
to the heavier quark mass points being beyond the radius of convergence of 
ChPT~\cite{Lin:2007pt,Bernard:2006te,RBC-UKQCD:2007}.

Instead of the SSE formula, we assume the following simple fit form, including 
the FVE in a way that respects the scaling observed in the data,
$
A + B m_\pi^2 + C f_V(m_\pi L)
$,
with $f_V(x) = \mathrm{e}^{-x}$, and
where $A, B,$ and $C$ are fit parameters.
We employ a constant term and one linear in the pion mass squared, the leading  contributions to $g_A$ in infinite volume.
The third term corresponds to the observed FVE,
taken as a function of $m_\pi L$ only, and vanishes rapidly 
towards the infinite volume limit, $L\to\infty$ at fixed pion mass.
The same $m_\pi L$ dependence appears in
one of the FVE contributions in Ref.~\cite{Jaffe:2001eb}.
We note that this simple form is used to estimate the FVE's
in the data but {\it not} the value of $g_A$ in the chiral limit
at fixed $L$. In the end, we chose this simplest form, in part, because
the fit result at the physical point is not sensitive to the particular choice of $f_V(x)$, as
discussed below.

In Fig.~\ref{fig:gagv_mpi} we see that
the 2+1 flavor data are described very well by
this simple fit ($\chi^2$/d.o.f$.=0.45$), using data computed on both volumes simultaneously.
The $L\to\infty$ extrapolation (solid line) in turn allows an extrapolation to the
physical pion mass ($m_\pi = 135$ MeV), $g_A = 1.20(6)(4)$,
where the first error is statistical.
The second error
is an estimate of the systematic error
determined by comparing this result with that from fits using different choices of $f_V(x)$, {\it e.g.},
the full form in~\cite{Jaffe:2001eb},
$x^{-3}$, and $m_\pi^2 \,\mathrm{e}^{-x} / x^{1/2}$. The latter is similar to HBChPT when 
$m_\pi L \gg 1$~\cite{Beane:2004rf,Detmold:2005pt,Khan:2006de}.
The extrapolated value is not
sensitive to the choice of $f_V$, and is also consistent
with a linear fit to the three heaviest points on the larger volume,
$g_A = 1.17(6)$.
We also fit our data, with and without the $f_V$ term,
to the 2-loop formula from HBChPT~\cite{Bernard:2006te} and find
that the extrapolated result is less than 1 and  
that the fits are generally unstable. This is because the many 
unknown low energy constants cannot be determined well from only 
four data points, even if some of them are fixed. More importantly, 
though the 2-loop formula extends the range of the chiral expansion, 
it is still only large enough to include our lightest point, as 
demonstrated in Ref.~\cite{Bernard:2006te}.
The systematic error arising from 
the difference of the renormalization constants for $A_\mu$
and $V_\mu$ is much smaller than the quoted systematic error.
From the fit result with $f_V(x)=\mathrm{e}^{-x}$, 
we estimate that if one aims to keep FVE's
at or below 1\%, then for $m_\pi =330$ MeV, spatial sizes of 3.4--4.1 fm 
$(m_\pi L\approx 5.7$--$6.9$) are necessary.

%

We have computed the nucleon axial charge with 2+1 flavors of  DWF on a largest volume
of $(2.7$ fm)$^3$,
a lightest $m_\pi$ of 330 MeV, and large statistics.
We have observed a downward $m_\pi$ dependence 
of $g_A$ on our larger volume,
and by comparing our results with those using different volumes, 
numbers of flavors, and lattice fermions as a function of the single variable $m_\pi L$,
concluded it is caused by the finite volume used in our calculation.
The data appear to scale well in this variable except in the quenched 
and partially quenched, mixed action
cases when the volume exceeds roughly (2.4 fm)$^3$,
which may be due to the presence of (partially-) quenched chiral logs.

We thank our colleagues in the RBC and UKQCD collaborations for helpful
discussions.
We thank Columbia University, University of Edinburgh, 
PPARC, RIKEN, BNL 
and the U.S.\ DOE for providing the QCDOC supercomputers used in this work.
T.B. and T.Y. were supported 
by the U.S.\ DOE under contract
DE-FG02-92ER40716. 
H.L. is supported by DOE contract DE-AC05-06OR23177 under which 
JSA, LLC operates THNAF.
S.S. is supported by JSPS (19540265).
J.Z. is supported by PPARC grant
PP/D000238/1.
\bibliography{letter_axch}

\end{document}